\documentclass[aps,prb,twocolumn,superscriptaddress,floatfix,10pt]{revtex4-2}

\usepackage{amsmath,amssymb,graphicx,bm,dsfont}
\usepackage{xcolor}
\usepackage[colorlinks=true,urlcolor=blue,linkcolor=blue,citecolor=blue,bookmarks=false]{hyperref}

\usepackage[english]{babel}
\usepackage{natbib}
\usepackage{ dsfont }
\usepackage{textcomp}
\usepackage{tikz}

\begin{document}
\title{Critical Sample Aspect Ratio and Magnetic Field Dependence for Antiskyrmion Formation in Mn\textsubscript{1.4}PtSn Single-Crystals}
\author{B. E. Zuniga Cespedes}
\affiliation{Institute of Applied Physics, Technische Universit{\"a}t Dresden, D-01069 Dresden, Germany}
\affiliation{Max Planck Institute for Chemical Physics of Solids, D-01187 Dresden, Germany}

\author{P. Vir}
\affiliation{Max Planck Institute for Chemical Physics of Solids, D-01187 Dresden, Germany}

\author{P. Milde}
\affiliation{Institute of Applied Physics, Technische Universit{\"a}t Dresden, D-01069 Dresden, Germany}

\author{C. Felser}
\affiliation{Max Planck Institute for Chemical Physics of Solids, D-01187 Dresden, Germany}
\affiliation{ct.qmat, Dresden-W\"{u}rzburg Cluster of Excellence-EXC 2147, TU Dresden, 01062 Dresden, Germany}

\author{L. M. Eng}
\affiliation{Institute of Applied Physics, Technische Universit{\"a}t Dresden, D-01069 Dresden, Germany}
\affiliation{ct.qmat, Dresden-W\"{u}rzburg Cluster of Excellence-EXC 2147, TU Dresden, 01062 Dresden, Germany}

\begin{abstract}
Mn\textsubscript{1.4}PtSn is the first material in which antiskyrmions have been observed in ultra-thin single crystalline specimens. While bulk crystals exhibit fractal patterns of purely ferromagnetic domain ordering at room temperature, ultra-thin Mn\textsubscript{1.4}PtSn lamellae clearly show antiskyrmion lattices with lattice spacings up to several ~\textmu m. In the work presented here, we systematically investigate the thickness region from 400~nm to 10~\textmu m using $100\times 100$~\textmu m\textsuperscript{2}-wide Mn\textsubscript{1.4}PtSn plates, and identify the critical thickness-to-width aspect ratio $\alpha_0 = 0.044$ for the ferromagnetic fractal domain to the non-collinear texture phase transition. Additionally, we also explore these non-collinear magnetic textures below the critical aspect ratio $\alpha_0$ above and below the spin-reorientation transition temperature $T$\textsubscript{SR} while applying variable external magnetic fields. What we find is a strong hysteresis for the occurrence of an antiskyrmion lattice, since the antiskyrmions preferentially nucleate by pinching them off from helical stripes in the transition to the field polarized state.
\end{abstract}

\maketitle

\section{Introduction}
Magnetic materials that exhibit topological spin arrangements are promising candidates for future applications, in particular for spintronics. One type of such topological arrangements are the so--called skyrmion lattices (SkLs) that may arise in the presence of an antisymmetric Dzyaloshinskii-Moriya interaction (DMI). Every skyrmion represents a spatial distribution of non-coplanarly arranged magnetic moments, whose orientations once cover all spatial directions. This mutual noncoplanar orientation of neighboring spins is generally described by the topological charge that, in the end, is able to differentiate between skyrmions (Skys) and antiskyrmions (A-Skys) \cite{koshibae2016,Huang2017,Camosi2018,Hoffmann2017,Zhang2017,Nagaosa2013}.

To date, a large number of skyrmion-hosting compounds has been discovered, such as the B20-type chiral magnets \cite{Muehlbauer2009,Muenzer2010,Yu:NatMat-2011-10-2,Wilhelm:PRL-2011-107-12,Milde2013}, $\beta$-Mn-type Co-Zn-Mn alloys~\cite{Tokunaga2015}, Cu$_2$OSeO$_3$~\cite{Seki2012,Adams2012, Zhang2016, Milde2016}, or some lacunar spinels \cite{Kezsmarki2015, Fujima:PRevB-2017-95-18, Bordacs:SR-2017-7-7584}, which all support Bloch- or N\'eel-type skyrmions, respectively. So far, antiskyrmions were observed in thin-plates of the tetragonal Heusler compounds Mn\textsubscript{1.4}Pt\textsubscript{0.9}Pd\textsubscript{0.1}Sn\cite{Nayak2017,Saha2019,Jena2020,Peng2020}, Mn\textsubscript{1.4}PtSn \cite{Ma2020}, Mn\textsubscript{1.3}Pt\textsubscript{1.0}Pd\textsubscript{0.1}Sn \cite{shimizu2020} and Mn\textsubscript{2}Rh\textsubscript{0.95}Ir\textsubscript{0.05}Sn \cite{Jena2020b}, only. These first Lorentz transmission electron microscopy (LTEM) measurements demonstrated the nucleation of a triangular antiskyrmion lattice (ASkL) in a magnetic field that was applied both perpendicular to the surface of a thin lamella sample, and parallel to the tetragonal $c$-axis of these crystals \cite{Nayak2017, Saha2019}.

Subsequent LTEM experiments revealed that for some particular temperature and field regions within the phase diagram as well as for selected sample thicknesses, A-Skys may also arrange in a squared lattice fashion ~\cite{Jena2020,Peng2020}. Moreover, elliptically-distorted skyrmions of both handedness as well as the non-topological bubble lattice were shown to appear when applying the symmetry-breaking in-plane magnetic field in combination with an out-of-plane field~\cite{Jena2020,Peng2020}. This makes the tetragonal Heusler compounds a unique class of materials hosting a rich variety of controllable magnetic textures. 

Moreover, in bulk samples of Mn\textsubscript{1.4}PtSn an anisotropic fractal ferromagnetic closure domain pattern (FMD) has been reported in clear contrast to the LTEM results, which shows, that Mn\textsubscript{1.4}PtSn is an easy-axis ferromagnet at room temperature with the easy-axis aligned along the tetragonal $c$-axis and a pronounced quadratic anisotropy within the $ab$-plane, and highlights the importance of long-ranged dipolar magnetic interactions for the formation of antiskyrmions in these materials \cite{sukhanov2020}. Also, a recent study of the magnetic textures in Mn\textsubscript{1.4}PtSn using wedge-shaped samples came to the same conclusion \cite{Ma2020}. Furthermore, at $T$\textsubscript{SR}$\sim$170K Mn\textsubscript{1.4}PtSn undergoes a spin reorientation transition from the coplanar above $T$\textsubscript{SR} to the non-coplanar spin orientation below $T$\textsubscript{SR} \cite{Vir2019a,Vir2019}.

Hence, to clearly determine the subtle differences in the phase diagram as well as to safely quantify the crossover from the reported fractal ferromagnetic domain (FMD) to the non-coplanar texture (NCT) phase in Mn\textsubscript{1.4}PtSn as needed for instance for realizing miniaturized room-temperature devices based on A-Skys, it is essential to characterize the different magnetic structures over a broad thickness range on samples with no or only weak thickness variations on the length scale of the antiskyrmion radius. We thus employ in the present study magnetic force microscopy (MFM) in order to clearly resolve the different magnetic textures of Mn\textsubscript{1.4}PtSn in Mn\textsubscript{1.4}PtSn plate samples cut out from bulk single crystals by means of focused ion beam (FIB) milling. What we observe is how the different magnetic patterns depend on sample thickness, and how and when these structures change when varying both temperature and the applied magnetic field.

Magnetic force microscopy (MFM) has proven to be a valuable tool when studying complex spin textures such as helices and skyrmions as well as complex domain patterns \cite{Milde2013,Milde2016,Kezsmarki2015,BacaniHug_2016, Hubert2008}. MFM can be effectuated with high precision not only under the various temperature and magnetic field conditions, but moreover, can be applied to investigate near-surface properties of any sample thickness, from monolayer to bulk systems. In this sense, MFM is not restricted to the analysis of ultra-thin lamellae, but will provide insight for any of these Mn\textsubscript{1.4}PtSn plate samples.

The paper is organized as follows: After describing the necessary experimental details, we (i) firstly discuss the thickness-dependent characteristic features of the real-space patterns obtained by MFM in zero field and above the spin-reorientation transition temperature $T$\textsubscript{SR}; here we clearly identify the critical thickness of $D_0 = 4.4$~\textmu m for Mn\textsubscript{1.4}PtSn plates, indicating the transition from the FMD to the NCT state. (ii) Secondly, we then focus on the field dependence and magnetic texture formation for a plate thickness of $D=2$~\textmu m well below the critical thickness $D_0$ at temperatures above $T$\textsubscript{SR}; of interest here are field-induced phase transitions between the two states. (iii) Finally, we briefly substantiate and discuss the impact of temperature, especially when cooling below  $T$\textsubscript{SR}.

\section{Experimental Details}
\begin{figure}
	\includegraphics[width=\columnwidth]{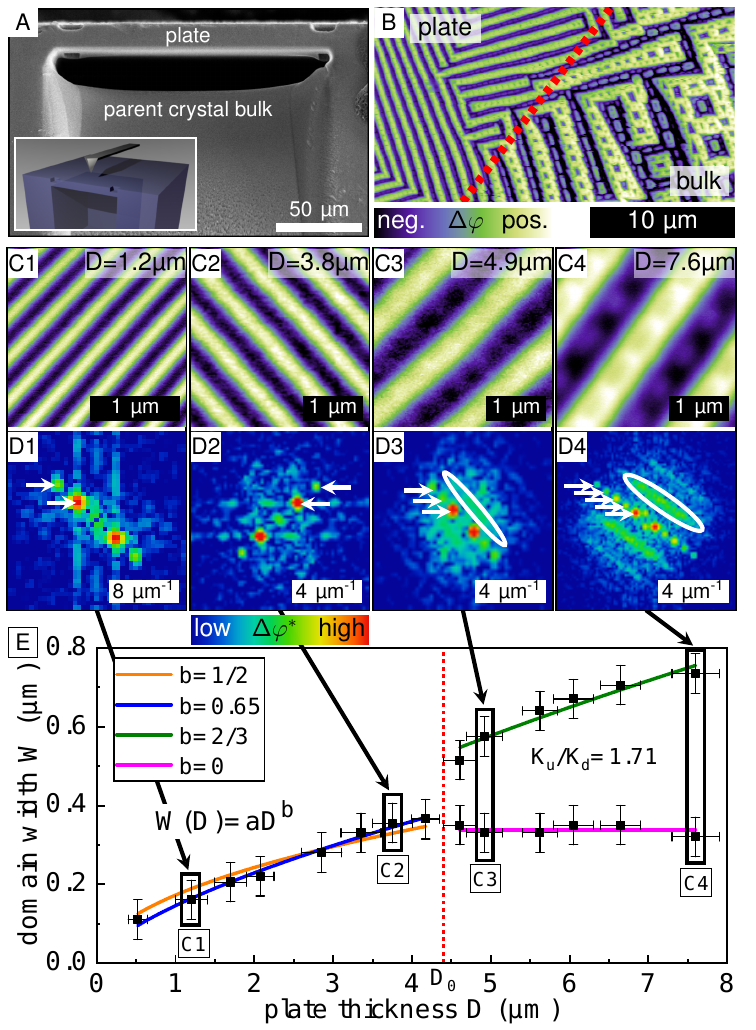}
	
	\caption{\label{fig1} Thickness dependency of the magnetic pattern in Mn\textsubscript{1.4}PtSn single crystal at 300K.
		(A) Side view of a thin-plate sample. The large surface is oriented normal to the $c$-axis. The inset schematically shows the geometry of the MFM measurement.
		(B) MFM image obtained at the end of a 10 \textmu m thick plate, where it is connected to the bulk (denoted by a red dotted line). The thickness dependent transformation of the magnetic pattern is clearly visible.
		(C1 to C4) MFM images obtained on plates with different thicknesses together with the corresponding FFT's plotted on logarithmic color scale (D1 to D4). Arrows highlight the peaks due to the stripe pattern. While in (C1) and (C2) the magnetic pattern is close to a sinusoidal wave in (C3) and (C4) the stripe pattern is closer to a square-wave with more anharmonicity. In the FFTs (D3 \& D4) stripes parallel to the peaks appear due to the nested domains (highlighted by ellipses). 
		(E) Domain width $W$ plotted as a function of plate thickness $D$. The first branching is identified at $D_0=4.4$~\textmu m. Individual branches can be fitted by a power law $W(D)=a \cdot D^b$.}	
\end{figure}

Single crystals of Mn\textsubscript{1.4}PtSn were grown by flux method using Sn as flux. For details please refer to Vir \textit{et.~al} \cite{Vir2019a}. 
After selecting crystals with appropriately oriented surfaces, a series of thin-plate samples with thicknesses ranging from $400$~nm to $10$~\textmu m suitable for MFM were FIB milled by cutting a $100\times 100$~\textmu m\textsuperscript{2} groove into the side of the single crystal using Xe ions at a current of 4 -- 60~nA. All samples hence are supported from the three (bulk) sides resulting in an extremely high mechanical stability for MFM inspection, as displayed in Fig.~\ref{fig1}(A). Note that all sample thicknesses were checked by scanning electron microscopy with an error of less than 50~nm. We also checked that gentle sample surface polishing using Xe or Ga ions at currents below 10~nA, did not impact the magnetic textures of any sample at all. 

MFM measurements were performed with two different instruments: Room-temperature inspection without applying any external magnetic fields was carried out on the Park Systems NX10~\cite{Park} using standard PPP-MFMR probes from Nanosensors~\cite{Nanosensors} at lift heights between 100 and 150~nm. Low-temperature measurements with external fields applied, were run in our Omicron cryogenic ultra-high vacuum STM/AFM instrument ~\cite{Omicron} operated with the R9 electronics from RHK~\cite{RHK}. Here we employed PPP-QMFMR probes from Nanosensors driven at mechanical oscillation amplitudes of $\approx 20$~nm at lift heights between 400 and 800~nm.

\section{Results}

\subsection{Thickness dependence}
The domain structure strongly depends on the sample thickness. In Fig.~\ref{fig1}(B) we show the magnetic pattern in zero field at room temperature when inspecting the border between the 10~\textmu m-thick Mn\textsubscript{1.4}PtSn plate and the bulk. While the bulk part exhibits the typical anisotropic fractal domain pattern, the magnetic pattern seen in the plate area is much simpler showing one row of nested domains, only. Note, however, that  magnetic domain walls in both areas are aligned along the same preferred crystallographic directions, i.e. along [100] and [010].

A series of MFM images obtained on thin-plates with different thicknesses is analyzed in Fig.~\ref{fig1}(C). Plates with thicknesses between 10~\textmu m and the treshhold value $D_0 = 4.4$~\textmu m clearly exhibit stripy patterns with one row of nested domains [see Figs.~\ref{fig1}(C3,C4)]. The corresponding 2-dimensional fast Fourier transforms (FFTs) (D3), (D4) display two distinct features, (i) a set of blurred streaks (highlighted by the white ellipse) originating from the nested domains, and (ii) a series of linearly aligned peaks within each streak (highlighted by arrows) that correspond to the higher Fourier components of the non-sinusoidal stripe profile. This behavior changes drastically for thicknesses below $D_0$, where the stripe profiles now become more sinusoidal showing no nested magnetic domains in real-space images. The FFTs in turn, show only a one dimensional pattern with less or no higher order peaks. Thus, $D_0$ clearly marks the critical thickness of the first branching, at which the magnetic pattern fundamentally changes.

From the MFM images we extract line profiles and measure the domain width $W$ or the helix period $\lambda = 2W$, respectively. For patterns with nested domains, we measure the width of both the base domain $W_b$ and of the smallest closure domains $W_s$. All measurements are summarized in Fig.~\ref{fig1}(E).

The dependence of magnetic domain structure on sample geometry in the case of thin infinitely extended plates has been described in detail for uniaxial ferromagnetic systems\cite{Hubert2008}. In general a power law dependence of the domain width $W$ on plate thickness $D$ is expected in the form $W(D)=a\cdot D^b$, where constants a and b are tied to the exchange constant $A$, the uniaxial anisotropy constant $K_u$ as well as the saturation magnetization $M_s$. For our experiment, the aspect ratio varies as a function of thickness, and therefore, we account for varying demagnetization factor $N$ resulting in a slight thickness dependency of factors a in our analysis. A more detailed summary can be found in the supplemental material.

In the thickness regime $D<D_0$, the domain width $W$ can be well fitted as a function of plate thickness $D$ by $W = a\cdot D^{1/2}$, as expected for stripe domains \cite{kittel1946, malek1958theory, kaplan1993domain} [orange line in Fig.~\ref{fig1}(E)]. Yet, leaving the exponent $\it{b}$ as a free fitting parameter $W\propto D^b$, we find $b=0.65\pm 0.03$, as shown in Fig.~\ref{fig1}(E) by the blue solid line. Since in this regime the observed texture is quite sinusoidal, the stray field is weaker than for uniform domains of the same width. In contrast to our observations, for only dipolar interactions this would result in a larger $W$ for constant $D$. The small deviations from the Kittel-law therefore may be attributed to the DMI that favors perpendicular spin arrangements and is responsible for the sinusoidal profile. 

For thicknesses $D>D_0$, free-exponent fitting for the base domain width $W_b$ results in $b = 0.66 \pm 0.08$ [green solid line in Fig.~\ref{fig1}(E)] in excellent agreement with the theoretical $b = 2/3$-behavior \cite{Hubert2008}. Meanwhile, the smallest surface domain width $W_s$ remains nearly constant as expected \cite{Hubert2008,lifshitz1992magnetic}. Thus, we conclude, that for thicknesses above $D_0$, Mn\textsubscript{1.4}PtSn is well described as a uniaxial ferromagnet with pronounced anisotropy within the $ab$-plane. Below $D_0$, DMI impacts dominantly such that spin helices instead of stripe domains nucleate.

From $D_0=4.4$~\textmu m, the saturation magnetization $M_s\approx4\times 10^5$A/m at room temperature\cite{Vir2019a}, and the ratio $K_u/K_d=1.71$ determined by the fit in the regime $D>D_0$ we find $K_u \approx 1.71 \cdot \mu_0 \cdot M_s^2/2 \approx 1.71\times10^{5}J/m^3$, which is small compared to common permanent magnets like Nd\textsubscript{2}Fe\textsubscript{14}B or SmCo\textsubscript{5} with $K_u=5\times10^6J/m^3$ and $K_u=1.7\times10^7J/m^3$ respectively \cite{Blundell2001} and on the same order of magnitude as the dipolar energy.

Thus, we may argue here, that magneto-crystalline and shape anisotropy compete and provide the effective, thickness-tunable $c$-axis anisotropy. While Mn\textsubscript{1.4}PtSn constitutes an easy-axis magnet along the $c$-axis, the shape anisotropy of a thin plate is thickness dependent and makes the surface normal of the extended surface to become the magnetic hard axis. For our plate samples investigated here, the $c$-axis is at the same time the surface normal, and hence the two anisotropies counteract. For cubic helimagnets with uniaxial anisotropy, it was predicted that stabilizing non-collinear spin textures might be possible for selected values of uniaxial anisotropies only \cite{Butenko2011, Roessler2010}. We thus may speculate that similarly the lowering of the effective anisotropy is the reason for the thickness-dependent fundamental change of the reported magnetic textures in Mn\textsubscript{1.4}PtSn. Therefore, instead of emphasizing on the relevance of an absolute thickness $D_0$, we may rather consider the critical thickness-to-width aspect ratio $\alpha_0$ for our quadratically-shaped plates as $\alpha_0 = D_0/100 \mu\textrm{m} = 0.044$. Nevertheless, we have to add that the special sample shape, where the plate is still connected to the bulk, may also have an impact on this value.

\begin{figure}
	\includegraphics[width=\columnwidth]{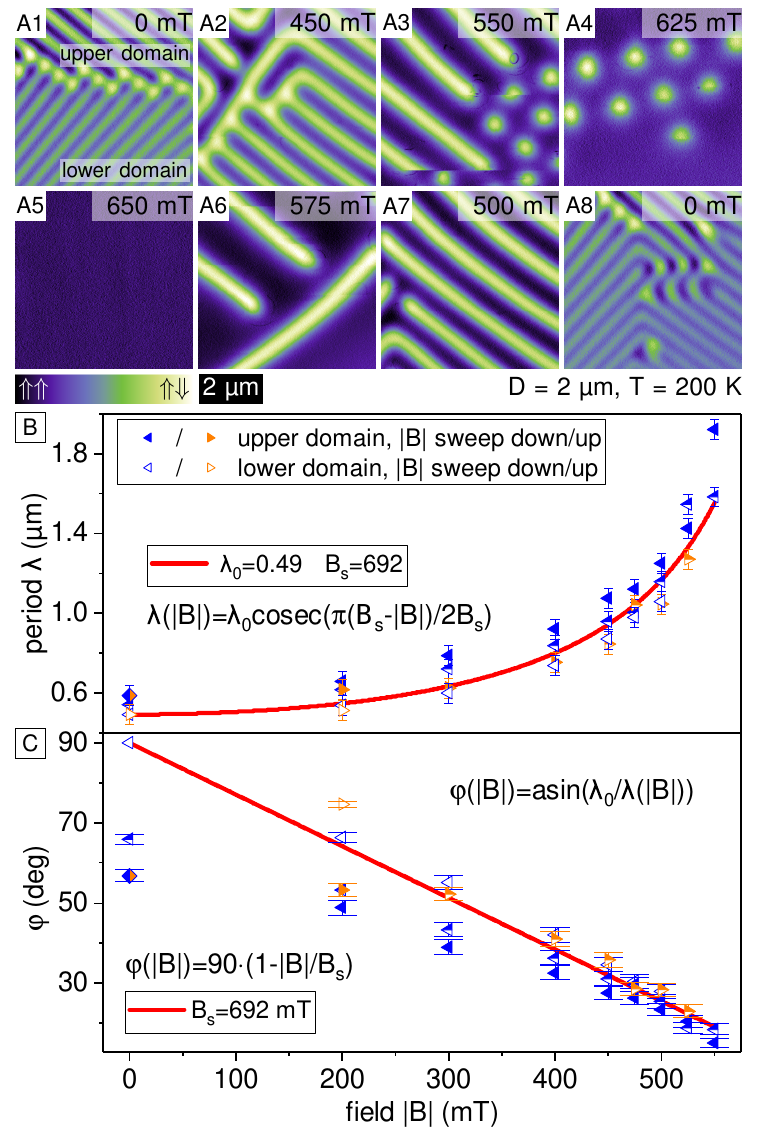}
	\caption{\label{fig2} Evolution of the magnetic pattern with field applied along the $c$-axis for a $D=2$~\textmu m thick plate at $T = 200$~K. 
	(A1 \& A2) Starting from zero field, a spin helix with field magnitude dependent period persists up to the critical field strength $B_{c1}=550$~mT.
	(A3) at $B_{c1}$ antiskyrmions start to pinch off from helical stripes eventually forming a hexagonal lattice (A4).
	(A5) Above $B_{c2} = 650$~mT the field polarized state is reached.
	(A6 to A8) For decreasing field strength the field polarized state persists until $B_{c2} = 600$~mT, where eventually single antiskyrmions nucleate as seeds for the helical stripe phase.
	(B)	The apparent helix period as function of magnetic field strength measured for increasing / decreasing field in blue / orange color, and for the upper / lower domain with filled / empty symbols, respectively.
	(C) The apparent period $\lambda$ can be interpreted as rotation of the wave-vector $\boldsymbol{q}$ into the field with a linear dependency $\phi(|B|)$.}
\end{figure}

\subsection{Field dependence}
Next, we focus on the field dependence and evolution of magnetic patterns for plates with thickness $D<D_0$ at $T>T_{SR}$, where ASkLs have been reported. The field dependent behavior in bulk samples has been discussed in a recent publication \cite{sukhanov2020}. A series of images measured for $D=2$~\textmu m at $T=200$~K is illustrated in Fig.~\ref{fig2}(A). The complete dataset  and datasets obtained at various other temperatures above $T_{SR}$ are available in the supplemental material.

Starting from zero external field as shown in Fig.~\ref{fig2}(A1), we find a spin helix up to the critical field strength $B_{c1}=550$~mT. With increasing field strength a systematic increase of the helix period $\lambda$ is observed [see Figs.~\ref{fig2}(A2\&A3)], closely resembling recent findings in Cu\textsubscript{2}OSeO\textsubscript{3} in the helical phase \cite{Milde2020}. Above $B_{c1}$, helical stripes break up into antiskyrmions [see Fig.~\ref{fig2}(A3)], that start to assemble in a hexagonal ASkL with a periodicity of approximately twice the helix period in zero field, as shown in Fig.~\ref{fig2}(A4). When further increasing the magnetic field, individual A-Skys start to unwind until the transition to the field-polarized state is reached at $B_{c2,up} = 650$~mT [see Fig.~\ref{fig2}(A5)]. In contrast for decreasing field strengths, the field-polarized state survives until $B_{c2,down} = 600$~mT, where eventually single A-Skys nucleate as seeds for the helical stripe phase [compare Fig.~\ref{fig2}(A6)]. Note, that no ASkL lattice is formed when lowering the external field from saturation, clearly indicating the hysteretic behavior at this phase transition. For $B<B_{c2}$, the helical phase is well stable down to zero magnetic field [see Figs.~\ref{fig2}(A7\&A8)]. Also note, that reversing the magnetic field excellently reproduces the above behavior, i.e. with  $B_{c2,up} = -650$~mT and $B_{c2,down} = -600$~mT. 

In Fig.~\ref{fig2}(B) we analyze the field dependence of the helix period $\lambda$. Apparently, the helix period changes as a function of external magnetic field for all domains. Moreover within the helical phase and as seen for instance in Fig.~\ref{fig2}(A1), we find magnetic domains within one and the same image that have different $\lambda$. In principle, there are two possible explanations for such a behavior:

Firstly, when assuming that the helix wave-vector $\boldsymbol{q}$ is always aligned in-plane, i.e. always lies in the image plane, then the helix transforms into an anharmonic helicoid and the magnitude of the period $\lambda$ must effectively change in magnetic field\cite{Butenko2011}. Yet, this explanation conflicts with our observations, as firstly domains with different period are clearly seen, and secondly the expected field dependency with an approximately constant period up to a critical field, where the period starts to diverge, does not fit to our data.

Secondly, a magnitude $\lambda$ that remains constant directly implies tilting of $\boldsymbol{q}$ with respect to the sample surface normal $\boldsymbol{n}$, i.e. the external magnetic field direction. What MFM then measures at the sample surface, is always the local projection onto the image plane, as has similarly been reported for the helical state in Cu\textsubscript{2}OSeO\textsubscript{3} \cite{Milde2020}. As a result, only the in-plane component of $\boldsymbol{q}$ causes the contrast in MFM. Thus, an apparently longer period $\lambda = 2\pi/[q\cdot\sin(\phi)] = \lambda_0/\sin(\phi)$ with $\phi = \sphericalangle(\boldsymbol{q},\boldsymbol{n})$ the inclination angle and $\lambda_0 = 2\pi/q$ the zero field period is measured. Vice versa, from the values of the apparent helix period $\lambda(B)$, we can compute $\phi(B) = \arcsin\left(\frac{\lambda_0}{\lambda(B)}\right)$. By doing so, we find an approximately linear dependence  $\phi(B)$ for the lower domain [see Fig. \ref{fig2}(C)]. Extrapolating this linear fit to $\phi(B_s)=0$ (i.e. the fully aligned case $\boldsymbol{q} || \boldsymbol{B}$) yields the critical field $B_s \approx 700$~mT, a value that is slightly larger than the onset of the field-polarized state deduced experimentally here. In contrast to Cu\textsubscript{2}OSeO\textsubscript{3}, the tilted helix becomes unstable already at an angle of $\approx 15${\textdegree }, long before $\boldsymbol{q} || \boldsymbol{B}$.

In small external fields, the upper domain deviates slightly from the linear behavior, which hints towards an incomplete $\boldsymbol{q}$-vector rotation, due to the very slow dynamics connected to the size of the domain and the large number of topological defects at the domain boundary. Similar long relaxation times were also reported for FeGe and MnSi \cite{Schoenherr2018, Bauer_2017}. In total, the field dependence of the apparent wavelength can be well explained by rotation of the helix wave-vector $\boldsymbol{q}$ into the external field with linear dependence $\varphi(B)$.

\subsection{Temperature dependence}

\begin{figure}
	\includegraphics[width=\columnwidth]{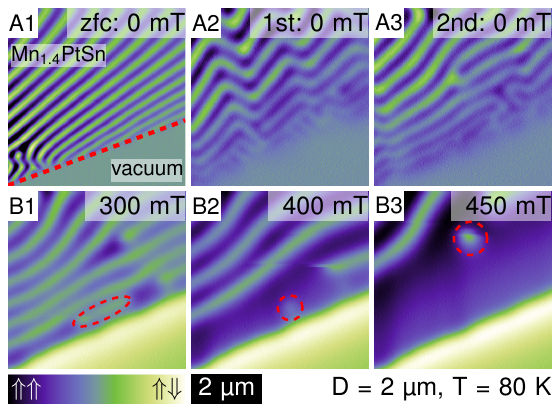}
	\caption{\label{fig3} Evolution of the magnetic pattern with field applied along the $c$-axis for a $D=2$~\textmu m thick plate at $T = 80$~K. 
		(A1 - A3) Comparing helix period and ordering after zero field cooling and after successive field sweeps up to saturation and back to zero field. The dashed line in (A1) marks the edge of the plate.
		(B1 - B3) In the transition to the field-polarized state only a few isolated bubbles can be spotted as remnants of stripe domains highlighted by dashed circles.}
\end{figure}

Finally, we extend our analysis to compare the magnetic pattern evolution in thin Mn\textsubscript{1.4}PtSn plates both above and below the spin-reorientation transition temperature $T$\textsubscript{SR} at 170~K. In bulk samples, for $T<T_{\text{SR}}$, the quadratic $ab$-plane anisotropy is lost, resulting in a much less rectangular domain pattern arrangement\cite{sukhanov2020}. 

When inspecting the plate with $D=2$ \textmu m, the effects on the magnetic pattern can be described as follows: Upon zero field cooling to $T=80$~K, the spin helices evolving above $T$\textsubscript{SR}, remain. Only when magnetically saturating the sample and ramping the field back to zero, can we notice some subtle changes in the magnetic pattern. In more detail, the period slightly increases and single phase fronts appear less straight, i.e. the helix wave-vector locally rotates slightly within the $ab$-plane [see Fig. S3(A1 to A3) and Fig. S1 in the Supplemental Material]. As for bulk samples, the quadratic $ab$-plane anisotropy characteristic above $T$\textsubscript{SR} vanishes. Even when ramping the field up into the field-polarized state showed no evidence of the occurrence of an ASkL, but only of single antiskyrmions or bubble domains [see Fig.3(B1 to B3)], in agreement with recent reports on Hall-effect measurements in bulk samples showing distinct behaviour above and below $T$\textsubscript{SR} \cite{Vir2019}.

\section{Discussion and conclusions}

In summary, we performed a thickness-dependent study of the magnetic patterns observed in Mn\textsubscript{1.4}PtSn, both under variable magnetic fields and over a wide temperature range. We found for our quadratic $100\times 100$~\textmu m\textsuperscript{2}-wide plates the critical thickness $D_0 = 4.4\pm0.4$~\textmu m corresponding to a critical thickness-to-width aspect ratio $\alpha_0 = 0.044\pm0.004$ above which the material can be well described as a uniaxial ferromagnet with a domain pattern fully built-up from stripe domains and nested closure domains.

Below $D_0$, a spin helix in low fields and an ASkL at fields close to saturation field were found, in agreement with literature. The phase diagram shows a strong hysteretic dependence, which manifests in the absence of the ASkL when the field is decreased from the saturated state. This might be connected to the observed A-Sky formation by pinching them off from helical stripes. Note that the reversed process had formerly been identified to unwind the SkL state towards the topological trivial helical state in Fe\textsubscript{0.5}Co\textsubscript{0.5}Si \cite{Milde2013}. The nucleation of further A-Skys in the field-polarized state happens to be less favorable than the area growth of already nucleated A-Skys into helical stripes, and hence, no ASkL but a stripe pattern forms. Thus, for selected magnetic field ranges, stable antiskyrmions may be easily nucleated by means of an external stimulus without formation of an ASkL at the same field and temperature. This is quite unusual for single crystalline systems, but offers the possibility to study single antiskyrmions at will in Mn\textsubscript{1.4}PtSn.

The apparent period of the spin helix shows a field dependence that can be well described by the rotation of the spin helix into the field direction, with a linear dependency of the inclination angle on the applied field. For temperatures below the spin-reorientation transition temperature the anisotropy within the $ab$-plane is lost, and the helix wave-vector may locally rotate within the $ab$-plane.

To conclude, in thin-plate samples of Mn\textsubscript{1.4}PtSn both the DMI as well as dipolar interactions are equally important for the stability of antiskyrmions. By varying the sample thickness and hence the effective $c$-axis anisotropy, the helical pitch and the ASkL size can be tuned to very large values as compared to other skyrmion hosting materials. Together with the likely ease to nucleate single antiskyrmions at will and at any predefined position by means of an external stimulus without formation of an ASkL, an even wider set of experimental techniques may be available such as optical detection with highest temporal resolution for studies of the antiskyrmion formation and dynamics.

\section*{Acknowledgments}

This project was funded by the German Research Foundation (DFG) under Grants No. EN~434/38-1 and MI~2004/3-1 as well as EN~434/40-1 as part of the Priority Program SPP 2137 ``Skyrmionics'', via the project C05 of the Collaborative Research Center SFB 1143 (project-id 247310070) at the TU Dresden and the W\"{u}rzburg-Dresden Cluster of Excellence on Complexity and Topology in Quantum Matter -- \textit{ct.qmat} (EXC 2147, project-id 390858490). B.E.Z.C. acknowledges support from the International Max Planck Research School for Chemistry and Physics of Quantum Materials (IMPRS-CPQM). 

\vfill

\bibliography{references}

\begin{thebibliography}{45}%
\makeatletter
\providecommand \@ifxundefined [1]{%
 \@ifx{#1\undefined}
}%
\providecommand \@ifnum [1]{%
 \ifnum #1\expandafter \@firstoftwo
 \else \expandafter \@secondoftwo
 \fi
}%
\providecommand \@ifx [1]{%
 \ifx #1\expandafter \@firstoftwo
 \else \expandafter \@secondoftwo
 \fi
}%
\providecommand \natexlab [1]{#1}%
\providecommand \enquote  [1]{``#1''}%
\providecommand \bibnamefont  [1]{#1}%
\providecommand \bibfnamefont [1]{#1}%
\providecommand \citenamefont [1]{#1}%
\providecommand \href@noop [0]{\@secondoftwo}%
\providecommand \href [0]{\begingroup \@sanitize@url \@href}%
\providecommand \@href[1]{\@@startlink{#1}\@@href}%
\providecommand \@@href[1]{\endgroup#1\@@endlink}%
\providecommand \@sanitize@url [0]{\catcode `\\12\catcode `\$12\catcode
  `\&12\catcode `\#12\catcode `\^12\catcode `\_12\catcode `\%12\relax}%
\providecommand \@@startlink[1]{}%
\providecommand \@@endlink[0]{}%
\providecommand \url  [0]{\begingroup\@sanitize@url \@url }%
\providecommand \@url [1]{\endgroup\@href {#1}{\urlprefix }}%
\providecommand \urlprefix  [0]{URL }%
\providecommand \Eprint [0]{\href }%
\providecommand \doibase [0]{https://doi.org/}%
\providecommand \selectlanguage [0]{\@gobble}%
\providecommand \bibinfo  [0]{\@secondoftwo}%
\providecommand \bibfield  [0]{\@secondoftwo}%
\providecommand \translation [1]{[#1]}%
\providecommand \BibitemOpen [0]{}%
\providecommand \bibitemStop [0]{}%
\providecommand \bibitemNoStop [0]{.\EOS\space}%
\providecommand \EOS [0]{\spacefactor3000\relax}%
\providecommand \BibitemShut  [1]{\csname bibitem#1\endcsname}%
\let\auto@bib@innerbib\@empty
\bibitem [{\citenamefont {Koshibae}\ and\ \citenamefont
  {Nagaosa}(2016)}]{koshibae2016}%
  \BibitemOpen
  \bibfield  {author} {\bibinfo {author} {\bibfnamefont {W.}~\bibnamefont
  {Koshibae}}\ and\ \bibinfo {author} {\bibfnamefont {N.}~\bibnamefont
  {Nagaosa}},\ }\bibfield  {title} {\bibinfo {title} {Theory of antiskyrmions
  in magnets},\ }\href@noop {} {\bibfield  {journal} {\bibinfo  {journal}
  {Nature communications}\ }\textbf {\bibinfo {volume} {7}},\ \bibinfo {pages}
  {10542} (\bibinfo {year} {2016})}\BibitemShut {NoStop}%
\bibitem [{\citenamefont {Huang}\ \emph {et~al.}(2017)\citenamefont {Huang},
  \citenamefont {Zhou}, \citenamefont {Chen}, \citenamefont {Shen},
  \citenamefont {Schmid}, \citenamefont {Liu},\ and\ \citenamefont
  {Wu}}]{Huang2017}%
  \BibitemOpen
  \bibfield  {author} {\bibinfo {author} {\bibfnamefont {S.}~\bibnamefont
  {Huang}}, \bibinfo {author} {\bibfnamefont {C.}~\bibnamefont {Zhou}},
  \bibinfo {author} {\bibfnamefont {G.}~\bibnamefont {Chen}}, \bibinfo {author}
  {\bibfnamefont {H.}~\bibnamefont {Shen}}, \bibinfo {author} {\bibfnamefont
  {A.~K.}\ \bibnamefont {Schmid}}, \bibinfo {author} {\bibfnamefont
  {K.}~\bibnamefont {Liu}},\ and\ \bibinfo {author} {\bibfnamefont
  {Y.}~\bibnamefont {Wu}},\ }\href@noop {} {\bibfield  {journal} {\bibinfo
  {journal} {Phys. Rev. B}\ }\textbf {\bibinfo {volume} {96}},\ \bibinfo
  {pages} {144412} (\bibinfo {year} {2017})}\BibitemShut {NoStop}%
\bibitem [{\citenamefont {Camosi}\ \emph {et~al.}(2018)\citenamefont {Camosi},
  \citenamefont {Rougemaille}, \citenamefont {Fruchart}, \citenamefont
  {Vogel},\ and\ \citenamefont {Rohart}}]{Camosi2018}%
  \BibitemOpen
  \bibfield  {author} {\bibinfo {author} {\bibfnamefont {L.}~\bibnamefont
  {Camosi}}, \bibinfo {author} {\bibfnamefont {N.}~\bibnamefont {Rougemaille}},
  \bibinfo {author} {\bibfnamefont {O.}~\bibnamefont {Fruchart}}, \bibinfo
  {author} {\bibfnamefont {J.}~\bibnamefont {Vogel}},\ and\ \bibinfo {author}
  {\bibfnamefont {S.}~\bibnamefont {Rohart}},\ }\href@noop {} {\bibfield
  {journal} {\bibinfo  {journal} {Phys. Rev. B}\ }\textbf {\bibinfo {volume}
  {97}},\ \bibinfo {pages} {134404} (\bibinfo {year} {2018})}\BibitemShut
  {NoStop}%
\bibitem [{\citenamefont {Hoffmann}\ \emph {et~al.}(2017)\citenamefont
  {Hoffmann}, \citenamefont {Zimmermann}, \citenamefont {M{\"u}ller},
  \citenamefont {Sch{\"u}rhoff}, \citenamefont {Kiselev}, \citenamefont
  {Melcher},\ and\ \citenamefont {Bl{\"u}gel}}]{Hoffmann2017}%
  \BibitemOpen
  \bibfield  {author} {\bibinfo {author} {\bibfnamefont {M.}~\bibnamefont
  {Hoffmann}}, \bibinfo {author} {\bibfnamefont {B.}~\bibnamefont
  {Zimmermann}}, \bibinfo {author} {\bibfnamefont {G.~P.}\ \bibnamefont
  {M{\"u}ller}}, \bibinfo {author} {\bibfnamefont {D.}~\bibnamefont
  {Sch{\"u}rhoff}}, \bibinfo {author} {\bibfnamefont {N.~S.}\ \bibnamefont
  {Kiselev}}, \bibinfo {author} {\bibfnamefont {C.}~\bibnamefont {Melcher}},\
  and\ \bibinfo {author} {\bibfnamefont {S.}~\bibnamefont {Bl{\"u}gel}},\
  }\href@noop {} {\bibfield  {journal} {\bibinfo  {journal} {Nat. Commun.}\
  }\textbf {\bibinfo {volume} {8}},\ \bibinfo {pages} {308} (\bibinfo {year}
  {2017})}\BibitemShut {NoStop}%
\bibitem [{\citenamefont {Zhang}\ \emph {et~al.}(2017)\citenamefont {Zhang},
  \citenamefont {Xia}, \citenamefont {Zhou}, \citenamefont {Liu}, \citenamefont
  {Zhang},\ and\ \citenamefont {Ezawa}}]{Zhang2017}%
  \BibitemOpen
  \bibfield  {author} {\bibinfo {author} {\bibfnamefont {X.}~\bibnamefont
  {Zhang}}, \bibinfo {author} {\bibfnamefont {J.}~\bibnamefont {Xia}}, \bibinfo
  {author} {\bibfnamefont {Y.}~\bibnamefont {Zhou}}, \bibinfo {author}
  {\bibfnamefont {X.}~\bibnamefont {Liu}}, \bibinfo {author} {\bibfnamefont
  {H.}~\bibnamefont {Zhang}},\ and\ \bibinfo {author} {\bibfnamefont
  {M.}~\bibnamefont {Ezawa}},\ }\href@noop {} {\bibfield  {journal} {\bibinfo
  {journal} {Nat. Commun.}\ }\textbf {\bibinfo {volume} {8}},\ \bibinfo {pages}
  {1717} (\bibinfo {year} {2017})}\BibitemShut {NoStop}%
\bibitem [{\citenamefont {Nagaosa}\ and\ \citenamefont
  {Tokura}(2013)}]{Nagaosa2013}%
  \BibitemOpen
  \bibfield  {author} {\bibinfo {author} {\bibfnamefont {N.}~\bibnamefont
  {Nagaosa}}\ and\ \bibinfo {author} {\bibfnamefont {Y.}~\bibnamefont
  {Tokura}},\ }\href@noop {} {\bibfield  {journal} {\bibinfo  {journal} {Nat.
  Nanotechnol.}\ }\textbf {\bibinfo {volume} {8}},\ \bibinfo {pages} {899}
  (\bibinfo {year} {2013})}\BibitemShut {NoStop}%
\bibitem [{\citenamefont {M\"uhlbauer}\ \emph {et~al.}(2009)\citenamefont
  {M\"uhlbauer}, \citenamefont {Binz}, \citenamefont {Jonietz}, \citenamefont
  {Pfleiderer}, \citenamefont {Rosch}, \citenamefont {Neubauer}, \citenamefont
  {Georgii},\ and\ \citenamefont {B\"oni}}]{Muehlbauer2009}%
  \BibitemOpen
  \bibfield  {author} {\bibinfo {author} {\bibfnamefont {S.}~\bibnamefont
  {M\"uhlbauer}}, \bibinfo {author} {\bibfnamefont {B.}~\bibnamefont {Binz}},
  \bibinfo {author} {\bibfnamefont {F.}~\bibnamefont {Jonietz}}, \bibinfo
  {author} {\bibfnamefont {C.}~\bibnamefont {Pfleiderer}}, \bibinfo {author}
  {\bibfnamefont {A.}~\bibnamefont {Rosch}}, \bibinfo {author} {\bibfnamefont
  {A.}~\bibnamefont {Neubauer}}, \bibinfo {author} {\bibfnamefont
  {R.}~\bibnamefont {Georgii}},\ and\ \bibinfo {author} {\bibfnamefont
  {P.}~\bibnamefont {B\"oni}},\ }\bibfield  {title} {\bibinfo {title} {Skyrmion
  lattice in a chiral magnet},\ }\href@noop {} {\bibfield  {journal} {\bibinfo
  {journal} {Science}\ }\textbf {\bibinfo {volume} {13}},\ \bibinfo {pages}
  {915} (\bibinfo {year} {2009})}\BibitemShut {NoStop}%
\bibitem [{\citenamefont {M\"unzer}\ \emph {et~al.}(2010)\citenamefont
  {M\"unzer}, \citenamefont {Neubauer}, \citenamefont {Adams}, \citenamefont
  {M\"uhlbauer}, \citenamefont {Franz}, \citenamefont {Jonietz}, \citenamefont
  {Georgii}, \citenamefont {B\"oni}, \citenamefont {Pedersen}, \citenamefont
  {Schmidt}, \citenamefont {Rosch},\ and\ \citenamefont
  {Pfleiderer}}]{Muenzer2010}%
  \BibitemOpen
  \bibfield  {author} {\bibinfo {author} {\bibfnamefont {W.}~\bibnamefont
  {M\"unzer}}, \bibinfo {author} {\bibfnamefont {A.}~\bibnamefont {Neubauer}},
  \bibinfo {author} {\bibfnamefont {T.}~\bibnamefont {Adams}}, \bibinfo
  {author} {\bibfnamefont {S.}~\bibnamefont {M\"uhlbauer}}, \bibinfo {author}
  {\bibfnamefont {C.}~\bibnamefont {Franz}}, \bibinfo {author} {\bibfnamefont
  {F.}~\bibnamefont {Jonietz}}, \bibinfo {author} {\bibfnamefont
  {R.}~\bibnamefont {Georgii}}, \bibinfo {author} {\bibfnamefont
  {P.}~\bibnamefont {B\"oni}}, \bibinfo {author} {\bibfnamefont
  {B.}~\bibnamefont {Pedersen}}, \bibinfo {author} {\bibfnamefont
  {M.}~\bibnamefont {Schmidt}}, \bibinfo {author} {\bibfnamefont
  {A.}~\bibnamefont {Rosch}},\ and\ \bibinfo {author} {\bibfnamefont
  {C.}~\bibnamefont {Pfleiderer}},\ }\bibfield  {title} {\bibinfo {title}
  {Skyrmion lattice in the doped semiconductor {F}e$_{1-x}${C}o$_x${S}i},\
  }\href@noop {} {\bibfield  {journal} {\bibinfo  {journal} {Phys. Rev. B}\
  }\textbf {\bibinfo {volume} {81}},\ \bibinfo {pages} {041203(R)} (\bibinfo
  {year} {2010})}\BibitemShut {NoStop}%
\bibitem [{\citenamefont {Yu}\ \emph {et~al.}(2011)\citenamefont {Yu},
  \citenamefont {Kanazawa}, \citenamefont {Onose}, \citenamefont {Kimoto},
  \citenamefont {Zhang}, \citenamefont {Ishiwata}, \citenamefont {Matsui},\
  and\ \citenamefont {Tokura}}]{Yu:NatMat-2011-10-2}%
  \BibitemOpen
  \bibfield  {author} {\bibinfo {author} {\bibfnamefont {X.}~\bibnamefont
  {Yu}}, \bibinfo {author} {\bibfnamefont {N.}~\bibnamefont {Kanazawa}},
  \bibinfo {author} {\bibfnamefont {Y.}~\bibnamefont {Onose}}, \bibinfo
  {author} {\bibfnamefont {K.}~\bibnamefont {Kimoto}}, \bibinfo {author}
  {\bibfnamefont {W.}~\bibnamefont {Zhang}}, \bibinfo {author} {\bibfnamefont
  {S.}~\bibnamefont {Ishiwata}}, \bibinfo {author} {\bibfnamefont
  {Y.}~\bibnamefont {Matsui}},\ and\ \bibinfo {author} {\bibfnamefont
  {Y.}~\bibnamefont {Tokura}},\ }\bibfield  {title} {\bibinfo {title} {Near
  room-temperature formation of a skyrmion crystal in thin-films of the
  helimagnet fege},\ }\href {https://doi.org/10.1038/NMAT2916} {\bibfield
  {journal} {\bibinfo  {journal} {Nature Materials}\ }\textbf {\bibinfo
  {volume} {10}},\ \bibinfo {pages} {106} (\bibinfo {year} {2011})}\BibitemShut
  {NoStop}%
\bibitem [{\citenamefont {Wilhelm}\ \emph {et~al.}(2011)\citenamefont
  {Wilhelm}, \citenamefont {Baenitz}, \citenamefont {Schmidt}, \citenamefont
  {R{\"o}{\ss}ler}, \citenamefont {Leonov},\ and\ \citenamefont
  {Bogdanov}}]{Wilhelm:PRL-2011-107-12}%
  \BibitemOpen
  \bibfield  {author} {\bibinfo {author} {\bibfnamefont {H.}~\bibnamefont
  {Wilhelm}}, \bibinfo {author} {\bibfnamefont {M.}~\bibnamefont {Baenitz}},
  \bibinfo {author} {\bibfnamefont {M.}~\bibnamefont {Schmidt}}, \bibinfo
  {author} {\bibfnamefont {U.}~\bibnamefont {R{\"o}{\ss}ler}}, \bibinfo
  {author} {\bibfnamefont {A.}~\bibnamefont {Leonov}},\ and\ \bibinfo {author}
  {\bibfnamefont {A.}~\bibnamefont {Bogdanov}},\ }\bibfield  {title} {\bibinfo
  {title} {Precursor phenomena at the magnetic ordering of the cubic helimagnet
  fege},\ }\href {https://doi.org/10.1103/PhysRevLett.107.127203} {\bibfield
  {journal} {\bibinfo  {journal} {Physical Review Letters}\ }\textbf {\bibinfo
  {volume} {107}},\ \bibinfo {pages} {127203} (\bibinfo {year}
  {2011})}\BibitemShut {NoStop}%
\bibitem [{\citenamefont {Milde}\ \emph {et~al.}(2013)\citenamefont {Milde},
  \citenamefont {K\"ohler}, \citenamefont {Seidel}, \citenamefont {Eng},
  \citenamefont {Bauer}, \citenamefont {Chacon}, \citenamefont {Kindervater},
  \citenamefont {M\"uhlbauer}, \citenamefont {Pfleiderer}, \citenamefont
  {Sch\"utte},\ and\ \citenamefont {Rosch}}]{Milde2013}%
  \BibitemOpen
  \bibfield  {author} {\bibinfo {author} {\bibfnamefont {P.}~\bibnamefont
  {Milde}}, \bibinfo {author} {\bibfnamefont {D.}~\bibnamefont {K\"ohler}},
  \bibinfo {author} {\bibfnamefont {J.}~\bibnamefont {Seidel}}, \bibinfo
  {author} {\bibfnamefont {L.}~\bibnamefont {Eng}}, \bibinfo {author}
  {\bibfnamefont {A.}~\bibnamefont {Bauer}}, \bibinfo {author} {\bibfnamefont
  {A.}~\bibnamefont {Chacon}}, \bibinfo {author} {\bibfnamefont
  {J.}~\bibnamefont {Kindervater}}, \bibinfo {author} {\bibfnamefont
  {S.}~\bibnamefont {M\"uhlbauer}}, \bibinfo {author} {\bibfnamefont
  {C.}~\bibnamefont {Pfleiderer}}, \bibinfo {author} {\bibfnamefont
  {C.}~\bibnamefont {Sch\"utte}},\ and\ \bibinfo {author} {\bibfnamefont
  {A.}~\bibnamefont {Rosch}},\ }\bibfield  {title} {\bibinfo {title} {Unwinding
  of a skyrmion lattice by magnetic monopoles},\ }\href@noop {} {\bibfield
  {journal} {\bibinfo  {journal} {Science}\ }\textbf {\bibinfo {volume}
  {340}},\ \bibinfo {pages} {1076} (\bibinfo {year} {2013})}\BibitemShut
  {NoStop}%
\bibitem [{\citenamefont {Tokunaga}\ \emph {et~al.}(2015)\citenamefont
  {Tokunaga}, \citenamefont {Yu}, \citenamefont {White}, \citenamefont
  {R\o{}nnow}, \citenamefont {Morikawa}, \citenamefont {Taguchi},\ and\
  \citenamefont {Tokura}}]{Tokunaga2015}%
  \BibitemOpen
  \bibfield  {author} {\bibinfo {author} {\bibfnamefont {Y.}~\bibnamefont
  {Tokunaga}}, \bibinfo {author} {\bibfnamefont {X.}~\bibnamefont {Yu}},
  \bibinfo {author} {\bibfnamefont {J.}~\bibnamefont {White}}, \bibinfo
  {author} {\bibfnamefont {H.}~\bibnamefont {R\o{}nnow}}, \bibinfo {author}
  {\bibfnamefont {D.}~\bibnamefont {Morikawa}}, \bibinfo {author}
  {\bibfnamefont {Y.}~\bibnamefont {Taguchi}},\ and\ \bibinfo {author}
  {\bibfnamefont {Y.}~\bibnamefont {Tokura}},\ }\bibfield  {title} {\bibinfo
  {title} {A new class of chiral materials hosting magnetic skyrmions beyond
  room temperature},\ }\href@noop {} {\bibfield  {journal} {\bibinfo  {journal}
  {Nat. Comm.}\ }\textbf {\bibinfo {volume} {6}},\ \bibinfo {pages} {7638}
  (\bibinfo {year} {2015})}\BibitemShut {NoStop}%
\bibitem [{\citenamefont {Seki}\ \emph {et~al.}(2012)\citenamefont {Seki},
  \citenamefont {Ishiwata},\ and\ \citenamefont {Tokura}}]{Seki2012}%
  \BibitemOpen
  \bibfield  {author} {\bibinfo {author} {\bibfnamefont {S.}~\bibnamefont
  {Seki}}, \bibinfo {author} {\bibfnamefont {S.}~\bibnamefont {Ishiwata}},\
  and\ \bibinfo {author} {\bibfnamefont {Y.}~\bibnamefont {Tokura}},\
  }\bibfield  {title} {\bibinfo {title} {Magnetoelectric nature of skyrmions in
  a chiral magnetic insulator {C}u$_{2}${OS}e{O}$_{3}$},\ }\href@noop {}
  {\bibfield  {journal} {\bibinfo  {journal} {Phys. Rev. B}\ }\textbf {\bibinfo
  {volume} {86}},\ \bibinfo {pages} {060403(R)} (\bibinfo {year}
  {2012})}\BibitemShut {NoStop}%
\bibitem [{\citenamefont {Adams}\ \emph {et~al.}(2012)\citenamefont {Adams},
  \citenamefont {Chacon}, \citenamefont {Wagner}, \citenamefont {Bauer},
  \citenamefont {Brandl}, \citenamefont {Pedersen}, \citenamefont {Berger},
  \citenamefont {Lemmens},\ and\ \citenamefont {Pfleiderer}}]{Adams2012}%
  \BibitemOpen
  \bibfield  {author} {\bibinfo {author} {\bibfnamefont {T.}~\bibnamefont
  {Adams}}, \bibinfo {author} {\bibfnamefont {A.}~\bibnamefont {Chacon}},
  \bibinfo {author} {\bibfnamefont {M.}~\bibnamefont {Wagner}}, \bibinfo
  {author} {\bibfnamefont {A.}~\bibnamefont {Bauer}}, \bibinfo {author}
  {\bibfnamefont {G.}~\bibnamefont {Brandl}}, \bibinfo {author} {\bibfnamefont
  {B.}~\bibnamefont {Pedersen}}, \bibinfo {author} {\bibfnamefont
  {H.}~\bibnamefont {Berger}}, \bibinfo {author} {\bibfnamefont
  {P.}~\bibnamefont {Lemmens}},\ and\ \bibinfo {author} {\bibfnamefont
  {C.}~\bibnamefont {Pfleiderer}},\ }\bibfield  {title} {\bibinfo {title}
  {Long-wavelength helimagnetic order and skyrmion lattice phase in
  {C}u$_{2}${OS}e{O}$_{3}$},\ }\href@noop {} {\bibfield  {journal} {\bibinfo
  {journal} {Phys. Rev. Lett.}\ }\textbf {\bibinfo {volume} {108}},\ \bibinfo
  {pages} {237204} (\bibinfo {year} {2012})}\BibitemShut {NoStop}%
\bibitem [{\citenamefont {Zhang}\ \emph {et~al.}(2016)\citenamefont {Zhang},
  \citenamefont {Bauer}, \citenamefont {Burn}, \citenamefont {Milde},
  \citenamefont {Neuber}, \citenamefont {Eng}, \citenamefont {Berger},
  \citenamefont {Pfleiderer}, \citenamefont {van~der Laan},\ and\ \citenamefont
  {Hesjedal}}]{Zhang2016}%
  \BibitemOpen
  \bibfield  {author} {\bibinfo {author} {\bibfnamefont {S.~L.}\ \bibnamefont
  {Zhang}}, \bibinfo {author} {\bibfnamefont {A.}~\bibnamefont {Bauer}},
  \bibinfo {author} {\bibfnamefont {D.~M.}\ \bibnamefont {Burn}}, \bibinfo
  {author} {\bibfnamefont {P.}~\bibnamefont {Milde}}, \bibinfo {author}
  {\bibfnamefont {E.}~\bibnamefont {Neuber}}, \bibinfo {author} {\bibfnamefont
  {L.~M.}\ \bibnamefont {Eng}}, \bibinfo {author} {\bibfnamefont
  {H.}~\bibnamefont {Berger}}, \bibinfo {author} {\bibfnamefont
  {C.}~\bibnamefont {Pfleiderer}}, \bibinfo {author} {\bibfnamefont
  {G.}~\bibnamefont {van~der Laan}},\ and\ \bibinfo {author} {\bibfnamefont
  {T.}~\bibnamefont {Hesjedal}},\ }\bibfield  {title} {\bibinfo {title}
  {Multidomain skyrmion lattice state in {C}u$_{2}${OS}e{O}$_{3}$},\
  }\href@noop {} {\bibfield  {journal} {\bibinfo  {journal} {Nano Letters}\
  }\textbf {\bibinfo {volume} {16}},\ \bibinfo {pages} {3285} (\bibinfo {year}
  {2016})}\BibitemShut {NoStop}%
\bibitem [{\citenamefont {Milde}\ \emph {et~al.}(2016)\citenamefont {Milde},
  \citenamefont {Neuber}, \citenamefont {Bauer}, \citenamefont {Pfleiderer},
  \citenamefont {Berger},\ and\ \citenamefont {Eng}}]{Milde2016}%
  \BibitemOpen
  \bibfield  {author} {\bibinfo {author} {\bibfnamefont {P.}~\bibnamefont
  {Milde}}, \bibinfo {author} {\bibfnamefont {E.}~\bibnamefont {Neuber}},
  \bibinfo {author} {\bibfnamefont {A.}~\bibnamefont {Bauer}}, \bibinfo
  {author} {\bibfnamefont {C.}~\bibnamefont {Pfleiderer}}, \bibinfo {author}
  {\bibfnamefont {H.}~\bibnamefont {Berger}},\ and\ \bibinfo {author}
  {\bibfnamefont {L.}~\bibnamefont {Eng}},\ }\bibfield  {title} {\bibinfo
  {title} {Heuristic description of magnetoelectricity of
  {C}u$_{2}${OS}e{O}$_{3}$},\ }\href@noop {} {\bibfield  {journal} {\bibinfo
  {journal} {Nano Lett.}\ }\textbf {\bibinfo {volume} {16}},\ \bibinfo {pages}
  {5612} (\bibinfo {year} {2016})}\BibitemShut {NoStop}%
\bibitem [{\citenamefont {K\'ezsm\'arki}\ \emph {et~al.}(2015)\citenamefont
  {K\'ezsm\'arki}, \citenamefont {Bord\'acs}, \citenamefont {Milde},
  \citenamefont {Neuber}, \citenamefont {Eng}, \citenamefont {White},
  \citenamefont {R\o{}nnow}, \citenamefont {Dewhurst}, \citenamefont
  {Mochizuki}, \citenamefont {Yanai}, \citenamefont {Nakamura}, \citenamefont
  {Ehlers}, \citenamefont {Tsurkan},\ and\ \citenamefont
  {Loidl}}]{Kezsmarki2015}%
  \BibitemOpen
  \bibfield  {author} {\bibinfo {author} {\bibfnamefont {I.}~\bibnamefont
  {K\'ezsm\'arki}}, \bibinfo {author} {\bibfnamefont {S.}~\bibnamefont
  {Bord\'acs}}, \bibinfo {author} {\bibfnamefont {P.}~\bibnamefont {Milde}},
  \bibinfo {author} {\bibfnamefont {E.}~\bibnamefont {Neuber}}, \bibinfo
  {author} {\bibfnamefont {L.}~\bibnamefont {Eng}}, \bibinfo {author}
  {\bibfnamefont {J.}~\bibnamefont {White}}, \bibinfo {author} {\bibfnamefont
  {H.}~\bibnamefont {R\o{}nnow}}, \bibinfo {author} {\bibfnamefont
  {C.}~\bibnamefont {Dewhurst}}, \bibinfo {author} {\bibfnamefont
  {M.}~\bibnamefont {Mochizuki}}, \bibinfo {author} {\bibfnamefont
  {K.}~\bibnamefont {Yanai}}, \bibinfo {author} {\bibfnamefont
  {H.}~\bibnamefont {Nakamura}}, \bibinfo {author} {\bibfnamefont
  {D.}~\bibnamefont {Ehlers}}, \bibinfo {author} {\bibfnamefont
  {V.}~\bibnamefont {Tsurkan}},\ and\ \bibinfo {author} {\bibfnamefont
  {A.}~\bibnamefont {Loidl}},\ }\bibfield  {title} {\bibinfo {title}
  {N\'el-type skyrmion lattice with confined orientation in the polar magnetic
  semiconductor gav$_{4}$s$_{8}$},\ }\href@noop {} {\bibfield  {journal}
  {\bibinfo  {journal} {Nat. Mat.}\ }\textbf {\bibinfo {volume} {14}},\
  \bibinfo {pages} {1116} (\bibinfo {year} {2015})}\BibitemShut {NoStop}%
\bibitem [{\citenamefont {Fujima}\ \emph {et~al.}(2017)\citenamefont {Fujima},
  \citenamefont {Abe}, \citenamefont {Tokunaga},\ and\ \citenamefont
  {Arima}}]{Fujima:PRevB-2017-95-18}%
  \BibitemOpen
  \bibfield  {author} {\bibinfo {author} {\bibfnamefont {Y.}~\bibnamefont
  {Fujima}}, \bibinfo {author} {\bibfnamefont {N.}~\bibnamefont {Abe}},
  \bibinfo {author} {\bibfnamefont {Y.}~\bibnamefont {Tokunaga}},\ and\
  \bibinfo {author} {\bibfnamefont {T.}~\bibnamefont {Arima}},\ }\bibfield
  {title} {\bibinfo {title} {Thermodynamically stable skyrmion lattice at low
  temperatures in a bulk crystal of lacunar spinel $gav_4se_8$},\ }\href
  {https://doi.org/10.1103/PhysRevB.95.180410} {\bibfield  {journal} {\bibinfo
  {journal} {Physical Review B}\ }\textbf {\bibinfo {volume} {95}},\ \bibinfo
  {pages} {180410(R)} (\bibinfo {year} {2017})}\BibitemShut {NoStop}%
\bibitem [{\citenamefont {Bord{\'a}cs}\ \emph {et~al.}(2017)\citenamefont
  {Bord{\'a}cs}, \citenamefont {Butykai}, \citenamefont {Szigeti},
  \citenamefont {White}, \citenamefont {Cubitt}, \citenamefont {Leonov},
  \citenamefont {Widmann}, \citenamefont {Ehlers}, \citenamefont {Krug~von
  Nidda}, \citenamefont {Tsurkan}, \citenamefont {Loidl},\ and\ \citenamefont
  {K{\'e}zsm{\'a}rki}}]{Bordacs:SR-2017-7-7584}%
  \BibitemOpen
  \bibfield  {author} {\bibinfo {author} {\bibfnamefont {S.}~\bibnamefont
  {Bord{\'a}cs}}, \bibinfo {author} {\bibfnamefont {{\'A}.}~\bibnamefont
  {Butykai}}, \bibinfo {author} {\bibfnamefont {B.}~\bibnamefont {Szigeti}},
  \bibinfo {author} {\bibfnamefont {J.}~\bibnamefont {White}}, \bibinfo
  {author} {\bibfnamefont {R.}~\bibnamefont {Cubitt}}, \bibinfo {author}
  {\bibfnamefont {A.}~\bibnamefont {Leonov}}, \bibinfo {author} {\bibfnamefont
  {S.}~\bibnamefont {Widmann}}, \bibinfo {author} {\bibfnamefont
  {D.}~\bibnamefont {Ehlers}}, \bibinfo {author} {\bibfnamefont {H.-A.}\
  \bibnamefont {Krug~von Nidda}}, \bibinfo {author} {\bibfnamefont
  {V.}~\bibnamefont {Tsurkan}}, \bibinfo {author} {\bibfnamefont
  {A.}~\bibnamefont {Loidl}},\ and\ \bibinfo {author} {\bibfnamefont
  {I.}~\bibnamefont {K{\'e}zsm{\'a}rki}},\ }\bibfield  {title} {\bibinfo
  {title} {Equilibrium skyrmion lattice ground state in a polar easy-plane
  magnet},\ }\href {https://doi.org/10.1038/s41598-017-07996-x} {\bibfield
  {journal} {\bibinfo  {journal} {Scientific Reports}\ }\textbf {\bibinfo
  {volume} {7}},\ \bibinfo {pages} {7584} (\bibinfo {year} {2017})}\BibitemShut
  {NoStop}%
\bibitem [{\citenamefont {Nayak}\ \emph {et~al.}(2017)\citenamefont {Nayak},
  \citenamefont {Kumar}, \citenamefont {Ma}, \citenamefont {Werner},
  \citenamefont {Pippel}, \citenamefont {Sahoo}, \citenamefont {Damay},
  \citenamefont {R{\"o}{\ss}ler}, \citenamefont {Felser},\ and\ \citenamefont
  {Parkin}}]{Nayak2017}%
  \BibitemOpen
  \bibfield  {author} {\bibinfo {author} {\bibfnamefont {A.~K.}\ \bibnamefont
  {Nayak}}, \bibinfo {author} {\bibfnamefont {V.}~\bibnamefont {Kumar}},
  \bibinfo {author} {\bibfnamefont {T.}~\bibnamefont {Ma}}, \bibinfo {author}
  {\bibfnamefont {P.}~\bibnamefont {Werner}}, \bibinfo {author} {\bibfnamefont
  {E.}~\bibnamefont {Pippel}}, \bibinfo {author} {\bibfnamefont
  {R.}~\bibnamefont {Sahoo}}, \bibinfo {author} {\bibfnamefont
  {F.}~\bibnamefont {Damay}}, \bibinfo {author} {\bibfnamefont {U.~K.}\
  \bibnamefont {R{\"o}{\ss}ler}}, \bibinfo {author} {\bibfnamefont
  {C.}~\bibnamefont {Felser}},\ and\ \bibinfo {author} {\bibfnamefont {S.~S.}\
  \bibnamefont {Parkin}},\ }\bibfield  {title} {\bibinfo {title} {Magnetic
  antiskyrmions above room temperature in tetragonal {Heusler} materials},\
  }\href@noop {} {\bibfield  {journal} {\bibinfo  {journal} {Nature}\ }\textbf
  {\bibinfo {volume} {548}},\ \bibinfo {pages} {561} (\bibinfo {year}
  {2017})}\BibitemShut {NoStop}%
\bibitem [{\citenamefont {Saha}\ \emph {et~al.}(2019)\citenamefont {Saha},
  \citenamefont {Srivastava}, \citenamefont {Ma}, \citenamefont {Jena},
  \citenamefont {Werner}, \citenamefont {Kumar}, \citenamefont {Felser},\ and\
  \citenamefont {Parkin}}]{Saha2019}%
  \BibitemOpen
  \bibfield  {author} {\bibinfo {author} {\bibfnamefont {R.}~\bibnamefont
  {Saha}}, \bibinfo {author} {\bibfnamefont {A.~K.}\ \bibnamefont
  {Srivastava}}, \bibinfo {author} {\bibfnamefont {T.}~\bibnamefont {Ma}},
  \bibinfo {author} {\bibfnamefont {J.}~\bibnamefont {Jena}}, \bibinfo {author}
  {\bibfnamefont {P.}~\bibnamefont {Werner}}, \bibinfo {author} {\bibfnamefont
  {V.}~\bibnamefont {Kumar}}, \bibinfo {author} {\bibfnamefont
  {C.}~\bibnamefont {Felser}},\ and\ \bibinfo {author} {\bibfnamefont
  {S.~S.~P.}\ \bibnamefont {Parkin}},\ }\href@noop {} {\bibfield  {journal}
  {\bibinfo  {journal} {Nat. Commun.}\ }\textbf {\bibinfo {volume} {10}},\
  \bibinfo {pages} {5305} (\bibinfo {year} {2019})}\BibitemShut {NoStop}%
\bibitem [{\citenamefont {Jena}\ \emph
  {et~al.}(2020{\natexlab{a}})\citenamefont {Jena}, \citenamefont {G{\"o}bel},
  \citenamefont {Ma}, \citenamefont {Kumar}, \citenamefont {Saha},
  \citenamefont {Mertig}, \citenamefont {Felser},\ and\ \citenamefont
  {Parkin}}]{Jena2020}%
  \BibitemOpen
  \bibfield  {author} {\bibinfo {author} {\bibfnamefont {J.}~\bibnamefont
  {Jena}}, \bibinfo {author} {\bibfnamefont {B.}~\bibnamefont {G{\"o}bel}},
  \bibinfo {author} {\bibfnamefont {T.}~\bibnamefont {Ma}}, \bibinfo {author}
  {\bibfnamefont {V.}~\bibnamefont {Kumar}}, \bibinfo {author} {\bibfnamefont
  {R.}~\bibnamefont {Saha}}, \bibinfo {author} {\bibfnamefont {I.}~\bibnamefont
  {Mertig}}, \bibinfo {author} {\bibfnamefont {C.}~\bibnamefont {Felser}},\
  and\ \bibinfo {author} {\bibfnamefont {S.~S.~P.}\ \bibnamefont {Parkin}},\
  }\href@noop {} {\bibfield  {journal} {\bibinfo  {journal} {Nat. Commun.}\
  }\textbf {\bibinfo {volume} {11}},\ \bibinfo {pages} {1115} (\bibinfo {year}
  {2020}{\natexlab{a}})}\BibitemShut {NoStop}%
\bibitem [{\citenamefont {Peng}\ \emph {et~al.}(2020)\citenamefont {Peng},
  \citenamefont {Takagi}, \citenamefont {Koshibae}, \citenamefont {Shibata},
  \citenamefont {Nakajima}, \citenamefont {Arima}, \citenamefont {Nagaosa},
  \citenamefont {Seki}, \citenamefont {Yu},\ and\ \citenamefont
  {Tokura}}]{Peng2020}%
  \BibitemOpen
  \bibfield  {author} {\bibinfo {author} {\bibfnamefont {L.}~\bibnamefont
  {Peng}}, \bibinfo {author} {\bibfnamefont {R.}~\bibnamefont {Takagi}},
  \bibinfo {author} {\bibfnamefont {W.}~\bibnamefont {Koshibae}}, \bibinfo
  {author} {\bibfnamefont {K.}~\bibnamefont {Shibata}}, \bibinfo {author}
  {\bibfnamefont {K.}~\bibnamefont {Nakajima}}, \bibinfo {author}
  {\bibfnamefont {T.}~\bibnamefont {Arima}}, \bibinfo {author} {\bibfnamefont
  {N.}~\bibnamefont {Nagaosa}}, \bibinfo {author} {\bibfnamefont
  {S.}~\bibnamefont {Seki}}, \bibinfo {author} {\bibfnamefont {X.}~\bibnamefont
  {Yu}},\ and\ \bibinfo {author} {\bibfnamefont {Y.}~\bibnamefont {Tokura}},\
  }\href@noop {} {\bibfield  {journal} {\bibinfo  {journal} {Nat.
  Nanotechnol.}\ }\textbf {\bibinfo {volume} {15}},\ \bibinfo {pages} {181}
  (\bibinfo {year} {2020})}\BibitemShut {NoStop}%
\bibitem [{\citenamefont {Ma}\ \emph {et~al.}(2020)\citenamefont {Ma},
  \citenamefont {Sharma}, \citenamefont {Saha}, \citenamefont {Srivastava},
  \citenamefont {Werner}, \citenamefont {Vir}, \citenamefont {Kumar},
  \citenamefont {Felser},\ and\ \citenamefont {Parkin}}]{Ma2020}%
  \BibitemOpen
  \bibfield  {author} {\bibinfo {author} {\bibfnamefont {T.}~\bibnamefont
  {Ma}}, \bibinfo {author} {\bibfnamefont {A.~K.}\ \bibnamefont {Sharma}},
  \bibinfo {author} {\bibfnamefont {R.}~\bibnamefont {Saha}}, \bibinfo {author}
  {\bibfnamefont {A.~K.}\ \bibnamefont {Srivastava}}, \bibinfo {author}
  {\bibfnamefont {P.}~\bibnamefont {Werner}}, \bibinfo {author} {\bibfnamefont
  {P.}~\bibnamefont {Vir}}, \bibinfo {author} {\bibfnamefont {V.}~\bibnamefont
  {Kumar}}, \bibinfo {author} {\bibfnamefont {C.}~\bibnamefont {Felser}},\ and\
  \bibinfo {author} {\bibfnamefont {S.~S.~P.}\ \bibnamefont {Parkin}},\
  }\href@noop {} {\bibfield  {journal} {\bibinfo  {journal} {Adv. Mater.}\
  }\textbf {\bibinfo {volume} {32}},\ \bibinfo {pages} {2002043} (\bibinfo
  {year} {2020})}\BibitemShut {NoStop}%
\bibitem [{\citenamefont {Shimizu}\ \emph {et~al.}(2020)\citenamefont
  {Shimizu}, \citenamefont {Nagase}, \citenamefont {So}, \citenamefont
  {Kuwahara}, \citenamefont {Ikarashi},\ and\ \citenamefont
  {Nagao}}]{shimizu2020}%
  \BibitemOpen
  \bibfield  {author} {\bibinfo {author} {\bibfnamefont {D.}~\bibnamefont
  {Shimizu}}, \bibinfo {author} {\bibfnamefont {T.}~\bibnamefont {Nagase}},
  \bibinfo {author} {\bibfnamefont {Y.-G.}\ \bibnamefont {So}}, \bibinfo
  {author} {\bibfnamefont {M.}~\bibnamefont {Kuwahara}}, \bibinfo {author}
  {\bibfnamefont {N.}~\bibnamefont {Ikarashi}},\ and\ \bibinfo {author}
  {\bibfnamefont {M.}~\bibnamefont {Nagao}},\ }\href@noop {} {\bibinfo {title}
  {Interaction between skyrmions and antiskyrmions in a coexisting phase of a
  {H}eusler material}} (\bibinfo {year} {2020}),\ \Eprint
  {https://arxiv.org/abs/arXiv:2008.07272} {arXiv:arXiv:2008.07272
  [cond-mat.mtrl-sci]} \BibitemShut {NoStop}%
\bibitem [{\citenamefont {Jena}\ \emph
  {et~al.}(2020{\natexlab{b}})\citenamefont {Jena}, \citenamefont {Stinshoff},
  \citenamefont {Saha}, \citenamefont {Srivastava}, \citenamefont {Ma},
  \citenamefont {Deniz}, \citenamefont {Werner}, \citenamefont {Felser},\ and\
  \citenamefont {Parkin}}]{Jena2020b}%
  \BibitemOpen
  \bibfield  {author} {\bibinfo {author} {\bibfnamefont {J.}~\bibnamefont
  {Jena}}, \bibinfo {author} {\bibfnamefont {R.}~\bibnamefont {Stinshoff}},
  \bibinfo {author} {\bibfnamefont {R.}~\bibnamefont {Saha}}, \bibinfo {author}
  {\bibfnamefont {A.~K.}\ \bibnamefont {Srivastava}}, \bibinfo {author}
  {\bibfnamefont {T.}~\bibnamefont {Ma}}, \bibinfo {author} {\bibfnamefont
  {H.}~\bibnamefont {Deniz}}, \bibinfo {author} {\bibfnamefont
  {P.}~\bibnamefont {Werner}}, \bibinfo {author} {\bibfnamefont
  {C.}~\bibnamefont {Felser}},\ and\ \bibinfo {author} {\bibfnamefont
  {S.~S.~P.}\ \bibnamefont {Parkin}},\ }\href@noop {} {\bibfield  {journal}
  {\bibinfo  {journal} {Nano Lett.}\ }\textbf {\bibinfo {volume} {20}},\
  \bibinfo {pages} {59} (\bibinfo {year} {2020}{\natexlab{b}})}\BibitemShut
  {NoStop}%
\bibitem [{\citenamefont {Sukhanov}\ \emph {et~al.}(2020)\citenamefont
  {Sukhanov}, \citenamefont {Zuniga~Cespedes}, \citenamefont {Vir},
  \citenamefont {Cameron}, \citenamefont {Heinemann}, \citenamefont {Martin},
  \citenamefont {Chaboussant}, \citenamefont {Kumar}, \citenamefont {Milde},
  \citenamefont {Eng}, \citenamefont {Felser},\ and\ \citenamefont
  {Inosov}}]{sukhanov2020}%
  \BibitemOpen
  \bibfield  {author} {\bibinfo {author} {\bibfnamefont {A.~S.}\ \bibnamefont
  {Sukhanov}}, \bibinfo {author} {\bibfnamefont {B.~E.}\ \bibnamefont
  {Zuniga~Cespedes}}, \bibinfo {author} {\bibfnamefont {P.}~\bibnamefont
  {Vir}}, \bibinfo {author} {\bibfnamefont {A.~S.}\ \bibnamefont {Cameron}},
  \bibinfo {author} {\bibfnamefont {A.}~\bibnamefont {Heinemann}}, \bibinfo
  {author} {\bibfnamefont {N.}~\bibnamefont {Martin}}, \bibinfo {author}
  {\bibfnamefont {G.}~\bibnamefont {Chaboussant}}, \bibinfo {author}
  {\bibfnamefont {V.}~\bibnamefont {Kumar}}, \bibinfo {author} {\bibfnamefont
  {P.}~\bibnamefont {Milde}}, \bibinfo {author} {\bibfnamefont {L.~M.}\
  \bibnamefont {Eng}}, \bibinfo {author} {\bibfnamefont {C.}~\bibnamefont
  {Felser}},\ and\ \bibinfo {author} {\bibfnamefont {D.~S.}\ \bibnamefont
  {Inosov}},\ }\bibfield  {title} {\bibinfo {title} {Anisotropic fractal
  magnetic domain pattern in bulk $\mathrm{Mn}_{1.4}\mathrm{Pt}\mathrm{Sn}$},\
  }\href {https://doi.org/10.1103/PhysRevB.102.174447} {\bibfield  {journal}
  {\bibinfo  {journal} {Phys. Rev. B}\ }\textbf {\bibinfo {volume} {102}},\
  \bibinfo {pages} {174447} (\bibinfo {year} {2020})}\BibitemShut {NoStop}%
\bibitem [{\citenamefont {Vir}\ \emph {et~al.}(2019{\natexlab{a}})\citenamefont
  {Vir}, \citenamefont {Kumar}, \citenamefont {Borrmann}, \citenamefont
  {Jamijansuren}, \citenamefont {Kreiner}, \citenamefont {Shekhar},\ and\
  \citenamefont {Felser}}]{Vir2019a}%
  \BibitemOpen
  \bibfield  {author} {\bibinfo {author} {\bibfnamefont {P.}~\bibnamefont
  {Vir}}, \bibinfo {author} {\bibfnamefont {N.}~\bibnamefont {Kumar}}, \bibinfo
  {author} {\bibfnamefont {H.}~\bibnamefont {Borrmann}}, \bibinfo {author}
  {\bibfnamefont {B.}~\bibnamefont {Jamijansuren}}, \bibinfo {author}
  {\bibfnamefont {G.}~\bibnamefont {Kreiner}}, \bibinfo {author} {\bibfnamefont
  {C.}~\bibnamefont {Shekhar}},\ and\ \bibinfo {author} {\bibfnamefont
  {C.}~\bibnamefont {Felser}},\ }\href@noop {} {\bibfield  {journal} {\bibinfo
  {journal} {Chem. Mater.}\ }\textbf {\bibinfo {volume} {31}},\ \bibinfo
  {pages} {5876} (\bibinfo {year} {2019}{\natexlab{a}})}\BibitemShut {NoStop}%
\bibitem [{\citenamefont {Vir}\ \emph {et~al.}(2019{\natexlab{b}})\citenamefont
  {Vir}, \citenamefont {Gayles}, \citenamefont {Sukhanov}, \citenamefont
  {Kumar}, \citenamefont {Damay}, \citenamefont {Sun}, \citenamefont
  {K{\"u}bler}, \citenamefont {Shekhar},\ and\ \citenamefont
  {Felser}}]{Vir2019}%
  \BibitemOpen
  \bibfield  {author} {\bibinfo {author} {\bibfnamefont {P.}~\bibnamefont
  {Vir}}, \bibinfo {author} {\bibfnamefont {J.}~\bibnamefont {Gayles}},
  \bibinfo {author} {\bibfnamefont {A.}~\bibnamefont {Sukhanov}}, \bibinfo
  {author} {\bibfnamefont {N.}~\bibnamefont {Kumar}}, \bibinfo {author}
  {\bibfnamefont {F.}~\bibnamefont {Damay}}, \bibinfo {author} {\bibfnamefont
  {Y.}~\bibnamefont {Sun}}, \bibinfo {author} {\bibfnamefont {J.}~\bibnamefont
  {K{\"u}bler}}, \bibinfo {author} {\bibfnamefont {C.}~\bibnamefont
  {Shekhar}},\ and\ \bibinfo {author} {\bibfnamefont {C.}~\bibnamefont
  {Felser}},\ }\bibfield  {title} {\bibinfo {title} {Anisotropic topological
  hall effect with real and momentum space berry curvature in the
  antiskrymion-hosting heusler compound {M}n$_{1.4}${P}t{S}n},\ }\href@noop {}
  {\bibfield  {journal} {\bibinfo  {journal} {Physical Review B}\ }\textbf
  {\bibinfo {volume} {99}},\ \bibinfo {pages} {140406(R)} (\bibinfo {year}
  {2019}{\natexlab{b}})}\BibitemShut {NoStop}%
\bibitem [{\citenamefont {Baćani}\ \emph {et~al.}(2016)\citenamefont
  {Baćani}, \citenamefont {Marioni}, \citenamefont {Schwenk},\ and\
  \citenamefont {Hug}}]{BacaniHug_2016}%
  \BibitemOpen
  \bibfield  {author} {\bibinfo {author} {\bibfnamefont {M.}~\bibnamefont
  {Baćani}}, \bibinfo {author} {\bibfnamefont {M.}~\bibnamefont {Marioni}},
  \bibinfo {author} {\bibfnamefont {J.}~\bibnamefont {Schwenk}},\ and\ \bibinfo
  {author} {\bibfnamefont {H.}~\bibnamefont {Hug}},\ }\bibfield  {title}
  {\bibinfo {title} {How to measure the local {D}zyaloshinskii-{M}oriya
  {I}nteraction in {S}kyrmion {T}hin {F}ilm {M}ultilayers},\ }\href
  {https://doi.org/10.1038/s41598-019-39501-x} {\bibfield  {journal} {\bibinfo
  {journal} {Scientific Reports}\ }\textbf {\bibinfo {volume} {9}} (\bibinfo
  {year} {2016})}\BibitemShut {NoStop}%
\bibitem [{\citenamefont {Hubert}\ and\ \citenamefont
  {Sch{\"a}fer}(2008)}]{Hubert2008}%
  \BibitemOpen
  \bibfield  {author} {\bibinfo {author} {\bibfnamefont {A.}~\bibnamefont
  {Hubert}}\ and\ \bibinfo {author} {\bibfnamefont {R.}~\bibnamefont
  {Sch{\"a}fer}},\ }\href@noop {} {\emph {\bibinfo {title} {Magnetic domains:
  the analysis of magnetic microstructures}}}\ (\bibinfo  {publisher} {Springer
  Science \& Business Media},\ \bibinfo {year} {2008})\BibitemShut {NoStop}%
\bibitem [{Par()}]{Park}%
  \BibitemOpen
  \href@noop {} {\bibinfo {title} {{P}ark {S}ystems {C}orp. {KANC} 15{F},
  {G}wanggyo-ro 109, {S}uwon 16229, {K}orea}}\BibitemShut {NoStop}%
\bibitem [{Nan()}]{Nanosensors}%
  \BibitemOpen
  \href@noop {} {\bibinfo {title} {{NANOSENSORS}\texttrademark, {R}ue
  {J}aquet-{D}roz 1, {C}ase {P}ostale 216, {CH}-2002 {N}euchatel,
  {S}witzerland}}\BibitemShut {NoStop}%
\bibitem [{Omi()}]{Omicron}%
  \BibitemOpen
  \href@noop {} {\bibinfo {title} {{O}micron {N}ano{T}echnology {G}mbh,
  {T}aunusstein, {G}ermany}}\BibitemShut {NoStop}%
\bibitem [{RHK()}]{RHK}%
  \BibitemOpen
  \href@noop {} {\bibinfo {title} {{RHK} {T}echnology, {I}nc., 1050 {E}ast
  {M}aple {R}oad, {T}roy, {MI} 48083 {USA}}}\BibitemShut {NoStop}%
\bibitem [{\citenamefont {Kittel}(1946)}]{kittel1946}%
  \BibitemOpen
  \bibfield  {author} {\bibinfo {author} {\bibfnamefont {C.}~\bibnamefont
  {Kittel}},\ }\bibfield  {title} {\bibinfo {title} {Theory of the structure of
  ferromagnetic domains in films and small particles},\ }\href@noop {}
  {\bibfield  {journal} {\bibinfo  {journal} {Physical Review}\ }\textbf
  {\bibinfo {volume} {70}},\ \bibinfo {pages} {965} (\bibinfo {year}
  {1946})}\BibitemShut {NoStop}%
\bibitem [{\citenamefont {M{\'a}lek}\ and\ \citenamefont
  {Kambersk{\`y}}(1958)}]{malek1958theory}%
  \BibitemOpen
  \bibfield  {author} {\bibinfo {author} {\bibfnamefont {Z.}~\bibnamefont
  {M{\'a}lek}}\ and\ \bibinfo {author} {\bibfnamefont {V.}~\bibnamefont
  {Kambersk{\`y}}},\ }\bibfield  {title} {\bibinfo {title} {On the theory of
  the domain structure of thin films of magnetically uni-axial materials},\
  }\href@noop {} {\bibfield  {journal} {\bibinfo  {journal} {Cechoslovackij
  fiziceskij zurnal}\ }\textbf {\bibinfo {volume} {8}},\ \bibinfo {pages} {416}
  (\bibinfo {year} {1958})}\BibitemShut {NoStop}%
\bibitem [{\citenamefont {Kaplan}\ and\ \citenamefont
  {Gehring}(1993)}]{kaplan1993domain}%
  \BibitemOpen
  \bibfield  {author} {\bibinfo {author} {\bibfnamefont {B.}~\bibnamefont
  {Kaplan}}\ and\ \bibinfo {author} {\bibfnamefont {G.}~\bibnamefont
  {Gehring}},\ }\bibfield  {title} {\bibinfo {title} {The domain structure in
  ultrathin magnetic films},\ }\href@noop {} {\bibfield  {journal} {\bibinfo
  {journal} {Journal of Magnetism and Magnetic Materials}\ }\textbf {\bibinfo
  {volume} {128}},\ \bibinfo {pages} {111} (\bibinfo {year}
  {1993})}\BibitemShut {NoStop}%
\bibitem [{\citenamefont {Lifshitz}\ \emph {et~al.}(1992)\citenamefont
  {Lifshitz}, \citenamefont {Ruderman} \emph {et~al.}}]{lifshitz1992magnetic}%
  \BibitemOpen
  \bibfield  {author} {\bibinfo {author} {\bibfnamefont {E.}~\bibnamefont
  {Lifshitz}}, \bibinfo {author} {\bibfnamefont {E.}~\bibnamefont {Ruderman}},
  \emph {et~al.},\ }\bibfield  {title} {\bibinfo {title} {On the magnetic
  structure of iron},\ }in\ \href@noop {} {\emph {\bibinfo {booktitle}
  {Perspectives in Theoretical Physics}}}\ (\bibinfo  {publisher} {Elsevier},\
  \bibinfo {year} {1992})\ pp.\ \bibinfo {pages} {203--218}\BibitemShut
  {NoStop}%
\bibitem [{\citenamefont {Blundell}(2001)}]{Blundell2001}%
  \BibitemOpen
  \bibfield  {author} {\bibinfo {author} {\bibfnamefont {S.}~\bibnamefont
  {Blundell}},\ }\href@noop {} {\emph {\bibinfo {title} {Magnetism in Condensed
  Matter}}}\ (\bibinfo  {publisher} {Oxford University Press},\ \bibinfo {year}
  {2001})\BibitemShut {NoStop}%
\bibitem [{\citenamefont {Butenko}\ \emph {et~al.}(2010)\citenamefont
  {Butenko}, \citenamefont {Leonov}, \citenamefont {R\"o\ss{}ler},\ and\
  \citenamefont {Bogdanov}}]{Butenko2011}%
  \BibitemOpen
  \bibfield  {author} {\bibinfo {author} {\bibfnamefont {A.~B.}\ \bibnamefont
  {Butenko}}, \bibinfo {author} {\bibfnamefont {A.~A.}\ \bibnamefont {Leonov}},
  \bibinfo {author} {\bibfnamefont {U.~K.}\ \bibnamefont {R\"o\ss{}ler}},\ and\
  \bibinfo {author} {\bibfnamefont {A.~N.}\ \bibnamefont {Bogdanov}},\
  }\bibfield  {title} {\bibinfo {title} {Stabilization of skyrmion textures by
  uniaxial distortions in noncentrosymmetric cubic helimagnets},\ }\href
  {https://doi.org/10.1103/PhysRevB.82.052403} {\bibfield  {journal} {\bibinfo
  {journal} {Phys. Rev. B}\ }\textbf {\bibinfo {volume} {82}},\ \bibinfo
  {pages} {052403} (\bibinfo {year} {2010})}\BibitemShut {NoStop}%
\bibitem [{\citenamefont {R{\"o}{\ss}ler}\ \emph {et~al.}(2011)\citenamefont
  {R{\"o}{\ss}ler}, \citenamefont {Leonov},\ and\ \citenamefont
  {Bogdanov}}]{Roessler2010}%
  \BibitemOpen
  \bibfield  {author} {\bibinfo {author} {\bibfnamefont {U.~K.}\ \bibnamefont
  {R{\"o}{\ss}ler}}, \bibinfo {author} {\bibfnamefont {A.~A.}\ \bibnamefont
  {Leonov}},\ and\ \bibinfo {author} {\bibfnamefont {A.~N.}\ \bibnamefont
  {Bogdanov}},\ }\bibfield  {title} {\bibinfo {title} {Chiral skyrmionic matter
  in non-centrosymmetric magnets},\ }\bibfield  {journal} {\bibinfo  {journal}
  {Journal of Physics: Conference Series}\ }\textbf {\bibinfo {volume} {303}},\
  \href {https://doi.org/10.1088/1742-6596/303/1/012105}
  {10.1088/1742-6596/303/1/012105} (\bibinfo {year} {2011})\BibitemShut
  {NoStop}%
\bibitem [{\citenamefont {Milde}\ \emph {et~al.}(2020)\citenamefont {Milde},
  \citenamefont {K\"ohler}, \citenamefont {Neuber}, \citenamefont {Ritzinger},
  \citenamefont {Garst}, \citenamefont {Bauer}, \citenamefont {Pfleiderer},
  \citenamefont {Berger},\ and\ \citenamefont {Eng}}]{Milde2020}%
  \BibitemOpen
  \bibfield  {author} {\bibinfo {author} {\bibfnamefont {P.}~\bibnamefont
  {Milde}}, \bibinfo {author} {\bibfnamefont {L.}~\bibnamefont {K\"ohler}},
  \bibinfo {author} {\bibfnamefont {E.}~\bibnamefont {Neuber}}, \bibinfo
  {author} {\bibfnamefont {P.}~\bibnamefont {Ritzinger}}, \bibinfo {author}
  {\bibfnamefont {M.}~\bibnamefont {Garst}}, \bibinfo {author} {\bibfnamefont
  {A.}~\bibnamefont {Bauer}}, \bibinfo {author} {\bibfnamefont
  {C.}~\bibnamefont {Pfleiderer}}, \bibinfo {author} {\bibfnamefont
  {H.}~\bibnamefont {Berger}},\ and\ \bibinfo {author} {\bibfnamefont {L.~M.}\
  \bibnamefont {Eng}},\ }\bibfield  {title} {\bibinfo {title} {Field-induced
  reorientation of helimagnetic order in ${\mathrm{cu}}_{2}{\mathrm{oseo}}_{3}$
  probed by magnetic force microscopy},\ }\href
  {https://doi.org/10.1103/PhysRevB.102.024426} {\bibfield  {journal} {\bibinfo
   {journal} {Phys. Rev. B}\ }\textbf {\bibinfo {volume} {102}},\ \bibinfo
  {pages} {024426} (\bibinfo {year} {2020})}\BibitemShut {NoStop}%
\bibitem [{\citenamefont {Schoenherr}\ \emph {et~al.}(2018)\citenamefont
  {Schoenherr}, \citenamefont {M\"uller}, \citenamefont {K\"ohler},
  \citenamefont {Rosch}, \citenamefont {Kanazawa}, \citenamefont {Tokura},
  \citenamefont {Garst},\ and\ \citenamefont {Meier}}]{Schoenherr2018}%
  \BibitemOpen
  \bibfield  {author} {\bibinfo {author} {\bibfnamefont {P.}~\bibnamefont
  {Schoenherr}}, \bibinfo {author} {\bibfnamefont {J.}~\bibnamefont
  {M\"uller}}, \bibinfo {author} {\bibfnamefont {L.}~\bibnamefont {K\"ohler}},
  \bibinfo {author} {\bibfnamefont {A.}~\bibnamefont {Rosch}}, \bibinfo
  {author} {\bibfnamefont {N.}~\bibnamefont {Kanazawa}}, \bibinfo {author}
  {\bibfnamefont {Y.}~\bibnamefont {Tokura}}, \bibinfo {author} {\bibfnamefont
  {M.}~\bibnamefont {Garst}},\ and\ \bibinfo {author} {\bibfnamefont
  {D.}~\bibnamefont {Meier}},\ }\bibfield  {title} {\bibinfo {title}
  {Topological domain walls in helimagnets},\ }\href@noop {} {\bibfield
  {journal} {\bibinfo  {journal} {Nat. Phys.}\ }\textbf {\bibinfo {volume}
  {14}},\ \bibinfo {pages} {465} (\bibinfo {year} {2018})}\BibitemShut
  {NoStop}%
\bibitem [{\citenamefont {Bauer}\ \emph {et~al.}(2017)\citenamefont {Bauer},
  \citenamefont {Chacon}, \citenamefont {Wagner}, \citenamefont {Halder},
  \citenamefont {Georgii}, \citenamefont {Rosch}, \citenamefont {Pfleiderer},\
  and\ \citenamefont {Garst}}]{Bauer_2017}%
  \BibitemOpen
  \bibfield  {author} {\bibinfo {author} {\bibfnamefont {A.}~\bibnamefont
  {Bauer}}, \bibinfo {author} {\bibfnamefont {A.}~\bibnamefont {Chacon}},
  \bibinfo {author} {\bibfnamefont {M.}~\bibnamefont {Wagner}}, \bibinfo
  {author} {\bibfnamefont {M.}~\bibnamefont {Halder}}, \bibinfo {author}
  {\bibfnamefont {R.}~\bibnamefont {Georgii}}, \bibinfo {author} {\bibfnamefont
  {A.}~\bibnamefont {Rosch}}, \bibinfo {author} {\bibfnamefont
  {C.}~\bibnamefont {Pfleiderer}},\ and\ \bibinfo {author} {\bibfnamefont
  {M.}~\bibnamefont {Garst}},\ }\bibfield  {title} {\bibinfo {title} {Symmetry
  breaking, slow relaxation dynamics, and topological defects at the
  field-induced helix reorientation in {M}n{S}i},\ }\href@noop {} {\bibfield
  {journal} {\bibinfo  {journal} {Phys. Rev. B}\ }\textbf {\bibinfo {volume}
  {95}},\ \bibinfo {pages} {024429} (\bibinfo {year} {2017})}\BibitemShut
  {NoStop}%
\end{thebibliography}%

\end{document}